\documentclass[acmsmall,nonacm]{acmart}

\usepackage[utf8]{inputenc}
\usepackage[T1]{fontenc}
\usepackage{booktabs}
\usepackage{url} 
\usepackage{graphicx} 
\usepackage{mathptmx} 
\usepackage{xspace}
\usepackage[binary-units=true]{siunitx} 
\usepackage{xcolor} 
\usepackage{listings} 
\usepackage{makecell} 
\usepackage{amsfonts}
\usepackage{blindtext} 
\usepackage{enumitem} 
\usepackage{tikz} 
\usepackage[export]{adjustbox} 
\usepackage[nolist]{acronym}
\usepackage{setspace} 
\usepackage[flushleft]{threeparttable} 
\usepackage[labelformat=simple]{subcaption} 
\usepackage{wrapfig}
\usepackage{tcolorbox} 
\usepackage{eurosym} 
\usepackage{pifont} 

\begin{acronym}[Bash]
\acro{IoT}[IoT]{Internet of Things}
\acro{MMM}[MMM]{memory map manager}
\acro{PCB}[PCB]{printed circuit board}
\acro{MCU}[MCU]{microcontroller unit}
\acrodefplural{MCU}[MCUs]{microcontroller units}
\acro{BPH}[BPH]{bluepill hat}
\acro{DUT}[DUT]{device-under-test}
\acrodefplural{DUT}[DUTs]{devices-under-test}
\acro{HAL}[HAL]{hardware abstraction layer}
\acrodefplural{HAL}[HALs]{hardware abstraction layers}
\acro{IRQ}[IRQ]{interrupt request}
\acro{OS}[OS]{operating system}
\acrodefplural{OS}[OSs]{operating systems}
\acro{MBT}[MBT]{model-based testing}

\end{acronym}

\DeclareSIUnit\baud{Bd}

\definecolor{mGreen}{rgb}{0,0.6,0}
\definecolor{dGray}{rgb}{0.8,0.8,0.8}
\definecolor{mPurple}{rgb}{0.58,0,0.82}
\definecolor{backgroundColour}{rgb}{0.95,0.95,0.92}
\lstdefinestyle{PyStyle}{
	backgroundcolor=\color{backgroundColour},
	commentstyle=\color{mGreen},
	keywordstyle=\color{magenta},
	otherkeywords={assert},
	numberstyle=\tiny\color{dGray},
	stringstyle=\color{mPurple},
	basicstyle=\footnotesize\ttfamily,
	breakatwhitespace=false,
	breaklines=true,
	captionpos=b,
	keepspaces=true,
	numbers=left,
	numbersep=5pt,
	showspaces=false,
	showstringspaces=false,
	showtabs=false,
	tabsize=2,
	language=Python
}

\definecolor{dgreen}{rgb}{0.0, 0.5, 0.0}
\newcommand{\cmark}{\textcolor{dgreen}{\ding{51}}}
\newcommand{\mcmark}{\textcolor{orange}{(\ding{51})}}
\newcommand{\xmark}{\textcolor{red}{\ding{55}}}

\newcommand{\eg}{\textit{e.g.,}~}
\newcommand{\ie}{\textit{i.e.,}~}
\newcommand{\etc}{\textit{etc.}~}
\newcommand{\one}{({\em i})\xspace}
\newcommand{\two}{({\em ii})\xspace}
\newcommand{\three}{({\em iii})\xspace}
\newcommand{\four}{({\em iv})\xspace}
\newcommand{\five}{({\em v})\xspace}

\newcommand{\tnode}{test node\xspace} 
\newcommand{\philip}{PHiLIP\xspace}

\newcommand{\mmm}{\ac{MMM}\xspace}

\newcommand{\philippal}{\texttt{philip\_pal}\xspace}
\newcommand{\etal}{\textit{et al.}~}

\makeatletter
\renewcommand{\paragraph}[1]{\vspace*{0.03in}\noindent{\bf #1.}\hspace{0.25ex \@plus1ex \@minus.2ex}}
\makeatother

\usetikzlibrary{
	arrows,
	arrows.meta,
	calc,
	colorbrewer,
	external,
	fit,
	matrix,
	positioning,
	shapes,
	shapes.misc,
	backgrounds,
	decorations.pathreplacing
}

\tikzstyle{line} = [draw, -latex']

\setcopyright{acmcopyright}
\acmJournal{TECS}
\acmYear{2021} \acmVolume{1} \acmNumber{1} \acmArticle{1}
\acmMonth{1} \acmPrice{15.00}\acmDOI{XX.YYYY/zzzzzzz}

\begin{document}
\title{\philip on the HiL: Automated Multi-platform OS Testing with External Reference Devices}
\author{Kevin Weiss}
\orcid{0000-0002-4786-2033}
\email{kevin.weiss@haw-hamburg.de}
\affiliation{%
	\institution{Hamburg University of Applied Sciences}
	\streetaddress{Berliner Tor 7}
	\city{Hamburg}
	\country{Germany}
	\postcode{20099}
}

\author{Michel Rottleuthner}
\orcid{0000-0003-2673-7679}
\email{michel.rottleuthner@haw-hamburg.de}
\affiliation{%
	\institution{Hamburg University of Applied Sciences}
	\streetaddress{Berliner Tor 7}
	\city{Hamburg}
	\country{Germany}
	\postcode{20099}
}

\author{Thomas C. Schmidt}
\orcid{0000-0002-0956-7885}
\email{t.schmidt@haw-hamburg.de}
\affiliation{%
	\institution{Hamburg University of Applied Sciences}
	\streetaddress{Berliner Tor 7}
	\city{Hamburg}
	\country{Germany}
	\postcode{20099}
}

\author{Matthias W\"ahlisch}
\orcid{0000-0002-3825-2807}
\email{m.waehlisch@fu-berlin.de}
\affiliation{%
	\institution{Freie Universit{\"a}t Berlin}
	\streetaddress{Takustr. 9}
	\city{Berlin}
	\country{Germany}
	\postcode{14195}
}



\renewcommand{\shortauthors}{Weiss, Rottleuthner, Schmidt, and W\"ahlisch}

\begin{abstract}
Developing an \ac{OS} for low-end embedded devices requires continuous adaptation to new hardware architectures and components, while serviceability of features needs to be assured for each individual platform under tight resource constraints.
It is challenging to design a versatile and accurate heterogeneous test environment that is agile enough to cover a continuous evolution of the code base and platforms.
This mission is even more challenging when organized in an agile open-source community process with many contributors such as for the RIOT OS.
Hardware in the Loop (HiL) testing and Continuous Integration (CI)  are automatable approaches to verify functionality, prevent regressions, and improve the overall quality at development speed in large community projects.

In this paper, we present \philip (Primitive Hardware in the Loop Integration Product), an open-source external reference device together with tools that validate the system software while it controls hardware and interprets physical signals.
Instead of focusing on a specific test setting, \philip takes the approach of a tool-assisted agile HiL test process, designed for continuous evolution and deployment cycles.
We explain its design, describe how it supports HiL tests, evaluate performance metrics, and report on practical experiences of employing \philip in an automated CI test infrastructure.
Our initial deployment comprises 22 unique platforms, each of which executes 98 peripheral tests every night.
\philip allows for easy extension of low-cost, adaptive testing infrastructures but serves testing techniques and tools to a much wider range of applications.

\end{abstract}

\maketitle

\definecolor{DarkGrey}{RGB}{37,37,37}
\definecolor{red}{RGB}{215,25,28}
\definecolor{orange}{RGB}{253,174,97}
\definecolor{yellow}{RGB}{255,255,191}
\definecolor{green}{RGB}{171,221,164}
\definecolor{philblue}{RGB}{63,161,226}
\definecolor{brightgray}{RGB}{255,255,179}
\definecolor{tnodesw}{RGB}{255,255,255}
\definecolor{config}{RGB}{255,255,255}

\colorlet{dutcolor}{yellow}
\colorlet{philipcolor}{philblue}
\colorlet{test}{brightgray}
\colorlet{testnodecolor}{green}
\colorlet{writecolor}{philipcolor}

\tikzset{
	testsuite/.pic = {
		\def\w{5.3ex};
		\def\h{6ex};
		\def\c{1ex};
		\def\r{1pt};
		\def\lw{1pt};
		\def\ln{6};
		\coordinate (nw) at (0,0);
		\coordinate (ne0) at ($(nw) + (\w, 0)$);
		\coordinate (ne1) at ($(ne0) - (\c, 0)$);
		\coordinate (ne2) at ($(ne0) - (0, \c)$);
		\coordinate (se) at ($(ne0) + (0, -\h)$);
		\coordinate (sw) at ($(nw) + (0, -\h)$);
		\filldraw [-, DarkGrey, line width = \lw, fill opacity=0] (nw) -- (ne1) -- (ne2)
				  [rounded corners=\r] -- (se) -- (nw|-se) -- (nw) -- cycle;
	    \draw [-, line width = \lw] (ne1) [rounded corners=\r]-- (ne1|-ne2) -- (ne2);
	    \node [anchor=north west, text width=4ex, text=DarkGrey] at ($(nw)+ (0,-.5ex)$)
	        {\setstretch{.66}\scriptsize\ttfamily\bfseries\tikzpictext\par};
        \foreach \k in {1,...,\ln}
        {
            \draw [-, DarkGrey, line width = \lw, line cap=round]
            ($(nw|-se) + (.5ex,.5ex) + (0,{(\k-1)/(\ln-1)*(\h - 4ex)})$)
            -- ++ ($(\w,0) - (1ex,0)$);
        }
        \begin{scope}
	        \node[fit=(nw) (se), inner sep = 0pt] (-all) {};
	    \end{scope}
    },
	firmware/.pic = {
		\def\w{5.3ex};
		\def\h{6ex};
		\def\c{1ex};
		\def\r{1pt};
		\def\lw{1pt};
		\def\ln{4};
		\coordinate (nw) at (0,0);
		\coordinate (ne0) at ($(nw) + (\w, 0)$);
		\coordinate (ne1) at ($(ne0) - (\c, 0)$);
		\coordinate (ne2) at ($(ne0) - (0, \c)$);
		\coordinate (se) at ($(ne0) + (0, -\h)$);
		\coordinate (sw) at ($(nw) + (0, -\h)$);
		\filldraw [-, DarkGrey, line width = \lw, fill=white] (nw) -- (ne1) -- (ne2)
				  [rounded corners=\r] -- (se) -- (nw|-se) -- (nw) -- cycle;
		\draw [-, line width = \lw] (ne1) [rounded corners=\r]-- (ne1|-ne2) -- (ne2);
		\node [anchor=north west, text width=4ex, text=DarkGrey] (text) at ($(nw)+ (0,-.5ex)$)
			{\setstretch{.66}\scriptsize\ttfamily\bfseries\tikzpictext\par};
		\node [text width=2ex] at ($(nw)+ (1.4ex,-2ex)$)
			{\setstretch{.5}\tiny\ttfamily 0101010 0010101 1010101 \par};
		\begin{scope}
		    \node[fit=(nw) (se), inner sep = 0pt] (-all) {};
		\end{scope}
	},
    componentmarker/.style={inner sep=2pt,circle,minimum width=.5ex,fill=none,draw,font=\LARGE\bfseries\sffamily },
}

\tikzstyle{write} = [rectangle, draw, fill=writecolor,
text width=3.4em, text centered, inner sep=2pt, font=\small]
\tikzstyle{execute} = [rectangle, draw, fill=white,
text centered, rounded corners, text width=3.3em, inner sep=2pt, font=\small]
\tikzstyle{action} = [draw, rounded corners,fill=execolor, text width=3.3em, inner sep=2pt, font=\small]

\section{Introduction}
\label{sec:intro}

The rapidly expanding \ac{IoT} faces the continuous arrival of new \acp{MCU}, peripherals, and platforms.
New and established components collectively comprise a zoo of embedded hardware platforms that admit various capabilities and a distinctively diverse set of features.
Application software is often requested to operate these devices with significant responsibility and high reliability. In this context, employing an embedded \ac{OS} with hardware abstraction can significantly ease development by making common code reusable and hardware independent.
High quality requirements on such system software, however, can only be ensured after extensive validation in realistic testing procedures.

Testing an embedded \ac{OS} can be challenging due to the constrained nature of the devices, the variety of hardware-specific behavior, and the requirements of real-world interactions.
As embedded devices interact with external hardware and the physical world, testing must also verify this behavior.
Rapidly evolving hardware and agile software development  require tests to be run regularly on a vast number of individual devices.
An automated, extensible way of testing a large variety of embedded devices is therefore required to develop and maintain a reliable embedded \ac{OS}.

\begin{figure}
\begin{subfigure}[t]{0.47\columnwidth}
\strut\vspace*{-\baselineskip}\newline
\begin{tikzpicture}
	\node[inner sep=0pt] (local-setup-photo) at (0,0)
	{\includegraphics[width=\columnwidth]{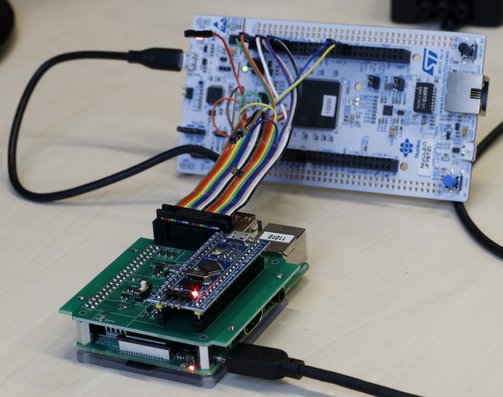}};
 	\draw node[componentmarker, fill=testnodecolor] (tnode-marker) at (-2.15,-1.52) {3};
 	\draw node[componentmarker, fill=philipcolor] (tnode-marker) at (-0.5,-0.7) {2};
 	\draw node[componentmarker, fill=dutcolor] (tnode-marker) at (1.3,0.9) {1};
\end{tikzpicture}
\vspace{0.8mm}
	\caption{A local HiL test setup that connects a device under test (DUT) (1) to \philip firmware in a bluepill board (2) which is mounted on a Raspberry Pi based \tnode (3).}
\label{fig:philip_to_dut}
\end{subfigure}
\hfill
\begin{subfigure}[t]{0.47\columnwidth}
\vspace{-3mm}
\begin{tikzpicture}
	\pic[draw,pic text = {TS A}] (ts_a) {testsuite};
	\node [color=black, below=-1.5ex of ts_a-all.south] (ts_dots) {\large\vdots};
	\pic[pic text = {TS N}, below=3ex of ts_a-all.south west] (ts_n) {testsuite};
	\node [fill=tnodesw, draw, rounded corners, right=5.5ex of ts_dots.east, text width=20ex, align=center] (test_logic) {\sffamily Test Framework};

	\draw[thick,decorate,decoration={brace,amplitude=2ex}] ([xshift=1ex]ts_a-all.north east) -- ([xshift=1ex]ts_n-all.south east) node [xshift=3ex,midway,rotate=90] (ts_brace) {};
	
	\pic[draw,pic text = {FW A}, below=2ex of ts_n-all.south west] (fw_a) {firmware};
	\node [color=black, below=-1.5ex of fw_a-all.south] (fw_dots) {\large\vdots};
	\pic[pic text = {FW N}, below=3ex of fw_a-all.south west] (fw_n) {firmware};
	\node [fill=dutcolor, draw, right=5ex of fw_dots.east, text width=8ex, align=center] (dut) {\sffamily DUT};
	\node [fill=philipcolor, draw, right=3ex of dut.east, text width=8ex, align=center] (philip) {\sffamily PHiLIP};
	\draw[thick,decorate,decoration={brace,amplitude=2ex}] ([xshift=1ex]fw_a-all.north east) -- ([xshift=1ex]fw_n-all.south east) node [xshift=3ex,midway,rotate=90] (fw_brace) {};
	
	\node [fill=tnodesw, draw, rounded corners, above=7ex of dut.north] (dut_pal) {\sffamily DUT PAL};
	\node [fill=tnodesw, draw, rounded corners, above=7ex of philip.north] (philip_pal) {\sffamily PHiLIP PAL};
	\node [fill=config, draw, rounded corners, right=2ex of philip_pal.east] (mmm) {\sffamily MMM};
	\pic[pic text = {Cfg.}, above=10ex of mmm.north, anchor=east, align=left, xshift=-2.6ex] (mmm_cfg) {testsuite};
	\pic[pic text = {\ FW}, below=12ex of mmm_cfg-all.south west] (fw_mmm) {firmware};
	
	\node [color=black, right=9.5ex of ts_ane2] {\sffamily Test Node};
	
	\path[draw,-, very thick] (dut) -- (dut_pal);
	\path[draw,-, very thick] (philip) -- (philip_pal);
	\path[draw,-, thick, dotted] (dut_pal) -- (test_logic);
	\path[draw,-, thick, dotted] (philip_pal) -- (test_logic);
	\path[draw,->, thick] (mmm) -- (philip_pal);
	\path[draw,->, thick] (mmm_cfg-all) -- (mmm);
	\path[draw,->, thick] (mmm) -- (fw_mmm-all);
	
	\draw[thick,decorate,decoration={brace,mirror,amplitude=1.25ex}] ([xshift=-0.8ex]fw_mmm-all.north west) -- ([xshift=-0.8ex]fw_mmm-all.south west) node [xshift=3ex,midway,rotate=90] (fw_brace) {};
	
	\coordinate (wire0) at ($(dut.south) + (0, -.5)$);
	\coordinate (wire1) at ($(philip.south) + (0, -.5)$);
	
	\path[draw,-, thick, dashed, double] (dut.south) -- (wire0)  -- node[pos=.5, below,font=\scriptsize\sffamily] {Peripheral Wiring} (wire1) -- (philip.south);
	
	\begin{scope}[on background layer]
		\node [draw, fill=testnodecolor, fit=(ts_a-all.north west) (philip_pal.south east)] {};
	\end{scope}
\end{tikzpicture}
\vspace{-5mm}
\caption{Overview of the HiL test environment: The \tnode runs test suites (TS) and interfaces with the DUT and \philip via Protocol Abstraction Layers (PAL). Each test suite corresponds to a test firmware (FW). \philip firmware and PAL are configured via the Memory Map Manager (MMM).}
\label{fig:Robot-FW-arch}
\end{subfigure}
\caption{Overview of physical components of the test setup and their architecture integration.}
\vspace{-3mm}
\end{figure}
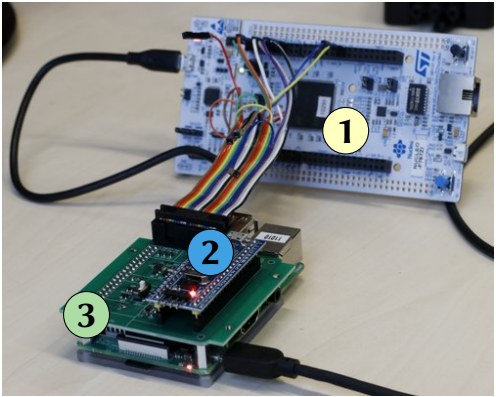
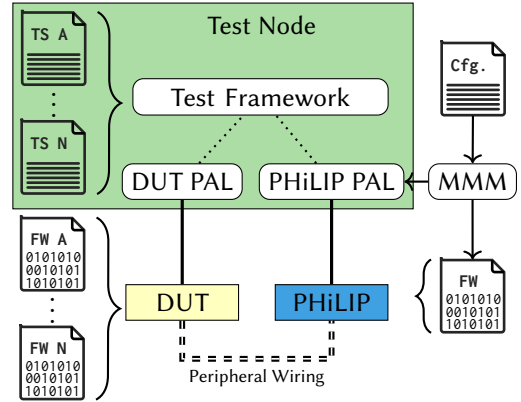

Many previous attempts in this area remained limited to individual use cases, without aiming for generality or focusing on a broad range of features and platforms.
Respectively, multi-platform Hardware in the Loop (HiL) testing is currently not covered well.
Moreover, previous solutions are often hard to acquire, set up, and maintain.
A versatile testing tool filling this gap should instead be compatible with a wide set of platforms while being easy to obtain, use, and extend.

In this paper, we try to bridge this gap between early research and reality for the open source OS RIOT~\cite{bhgws-rotoi-13}.
We propose a layered testing architecture that enables agile test development and employs an external reference device to provide automated HiL testing for a large variety of embedded devices.
Following this approach, we intend to close the loop from research to design, engineering, and further to operations, triggering a rich set of feedback.
Lessons learned substantiated research and design work during the years of developing and optimizing \philip.
A strong interaction with the large RIOT community was part of this process.

The concept considers three main components.
One is a resource-constrained \ac{DUT} that is evaluated (see \autoref{fig:philip_to_dut}).
The second is a reference device that takes measurements and executes hardware interactions with the \ac{DUT}.
The third referred to as \textit{\tnode}, coordinates, and controls the two previously introduced entities.
To streamline the integration and maintenance of implemented tools and firmware, our concept also includes the \mmm~that simplifies versioning and coordination of configuration data and documentation across device and tool boundaries.
This is shown in \autoref{fig:Robot-FW-arch} and will be explained in \autoref{sec:sys}.
In particular, this paper makes the following key contributions:
\begin{enumerate}[noitemsep]
	\item Following an in depth requirements analysis, we design a testing abstraction layer and a structured testing interface.
	\item We introduce \philip, our Primitive HiL Integration Product, as a firmware with verified peripheral behavior together with a tool-set for agile test development.
	\item We integrate \philip with multi-platform \acp{DUT} into a fully automated HiL testing environment and report on lessons learned based on our deployment.
	\item We evaluate our HiL testing proposal from the perspectives of testing impact, resource expenditures, and scalability.
\end{enumerate}

About half of the current work on IoT testing concerns interoperability and testbeds~\cite{abfc-aqits-19}.
Interoperability testing mainly targets networking~\cite{mkmay-itda-19}, often more narrowly wireless protocol conformance~\cite{lzgmw-tsrwp-18}, or protocol performance~\cite{gklpf-inpmm-21}.
Device heterogeneity is still a major challenge~\cite{tessl-gfky-18}, though.
Only very few contributions exist that are validated against real open-source embedded software, whereas the majority uses simplified examples.
The fast-paced business culture and lacking standardization in this area~\cite{t-seclm-18} leads to sidelining of serious challenges regarding product quality, security, and privacy~\cite{gb-ctcit-19}.
Together with the prevalent resource constraints, devices are often shipped without complex code that prevents or corrects errors at run time---eventually leading to faulty behavior and reliability problems of applications~\cite{slas-twtmpw-19}.
All those observations strongly support the relevance of our multi-platform testing for hardware layer abstractions of a popular open-source \ac{OS}.

In the remainder of this paper, we present \philip together with its design concepts, functions, tools, and experiences from our long-term deployment in the wild.
The presentation is organized as follows.
\autoref{sec:problem} discusses background and challenges related to testing embedded devices and \acp{OS}.
An overview of our proposed HiL testing architecture is given in \autoref{sec:sys}.
\autoref{sec:philip} dives into the details of \philip, a key component of our solution.
\autoref{sec:tests} explains how our setup is used for automated multi-platform testing of \ac{OS} hardware abstraction modules.
We evaluate our approach in \autoref{sec:evaluation} and report on its key performance metrics.
Other work is related in \autoref{sec:related_work}.
Some lessons learned and potential improvements are discussed in \autoref{sec:discussion} together with conclusions and an outlook on future work.

\section{Testing Embedded Systems: Challenges and Requirements}
\label{sec:problem}

Software testing commonly consists of functional~\cite{r-sutp-06} and non-functional tests~\cite{mkm-tnfra-07}.
Methodologies such as test-driven development~\cite{kwbf-mewtd-07} or \ac{MBT}~\cite{kr-mbtes-12} can be applied at almost any stage of the development process.
Even though it seems suitable to approach testing of IoT solutions via conventional software testing levels~\cite{tc-stlit-19}, embedded systems at the lower end of IoT architectures put up a unique set of nontrivial challenges.
In particular, their inherent technical properties and requirements, their heterogeneity and hardware proximity must be taken into account~\cite{bga-itccq-18, dcpf-boett-18, abfc-aqits-19}.
Even at its core, IoT systems comprise very specific hardware peripherals, among which are timers~\cite{grs-wltha-21},  energy management components~\cite{rsw-sypea-21}, entropy sources~\cite{ksw-gpngi-21}, and crypto chips~\cite{kblsw-pscli-21}.

The low quality of IoT system tests indicates that existing testing approaches still face difficulties overcoming these domain-specific challenges~\cite{cssms-astit-19}.
Existing gaps are partially attributed to a mismatched focus between industry and academia~\cite{dcpf-boett-18}.
Falling in line with others that identified the need for joined work between academia and industry to enable new testing approaches in real environments~\cite{cssms-astit-19}.

\subsection{Multi-platform Testing Needs Automation}

Developers of IoT systems can easily access various hardware features and specific functions when relying on the \ac{HAL} and APIs of an \ac{OS} for embedded devices.
This flexibility for the application developer comes at a cost for the \ac{OS} developers.
Each device-specific implementation must be maintained and thoroughly tested.
Whenever a feature is added or an API is changed, all implementations must be tested and validated again.
Overall, testing often accounts for more than 50 percent of development costs~\cite{abccc-osmas-13}, providing a very strong incentive to automate testing steps.

Several popular IoT \acp{OS} are open source and driven by a diverse and distributed community of users and developers who report problems, provide new features, and fix bugs.
It is therefore imperative for these projects to automate testing and verification processes for community contributions.
Many other popular embedded \acp{OS} such as RIOT~\cite{bghkl-rosos-18}, Contiki~\cite{dgv-clfos-04}, Mbed~\cite{arm-mbed-20}, and Zephyr~\cite{zephyr-20} include testing infrastructure that provides static and unit tests.
Some of these \acp{OS} have limited board simulation support but are not included in automated testing.
RIOT, Mbed, and Zephyr support HiL testing on a subset of boards but struggle with heterogeneous multi-platform testing with different architectures and boards.

Aligned to this context, the goal for our automated multi-platform testing involves two major parts.
First, an architecture is needed that enables testing unified \ac{OS} interfaces for the same deterministic behavior across all supported hardware.
This includes correct program flows, handling of parameters, API return codes, internal state, and acquire/release operations for resources.
Second, the verification of physical IO signals to comply with our specification and involved communication standards, \ie interactions to the physical world and other devices.
Most importantly this involves the need for a reference device to instrument our \acp{DUT} in a generic way.

\subsection{Agile and Reproducible HiL Testing}

HiL setups are commonly realized by connecting the \ac{DUT} to external hardware, which represents subsystems that are part of the final product or its environment.
A motor control \ac{DUT} could, for example, be connected to a real motor or the circuit that powers the motor.
In our scenario of \ac{OS} development, the general-purpose \acp{DUT} do not have such a single specific use case.
Therefore, we simply call the external subsystem that pairs with our \ac{DUT} \textit{reference device}.

\begin{wraptable}{R}{8cm}
	\setlength{\tabcolsep}{0.4em}
	\caption{
		Feature comparison of solutions for external test devices: Logic Analyzer (LA), Emulator/Simulator~(ES), Slave Device (SD), or FPGA versus commodity MCU.
	}
	\label{table:test_device_solutions}
	\begin{tabular}{ lccccc }
		\hline
		{\bf Metric} & {\bf LA} & {\bf ES} & {\bf SD} & {\bf FPGA} & {\bf MCU}\\
		\hline
		{\bf HW Control }		& \mcmark & \mcmark & \mcmark & \cmark & \cmark \\
		{\bf Reproducible Setup}	& \cmark & \cmark & \cmark & \mcmark & \cmark \\
		{\bf Agile Adaptability}		& \xmark & \xmark & \xmark & \mcmark & \cmark\\
		{\bf Multi-Platform}	& \cmark & \xmark & \cmark & \cmark & \cmark \\
		\hline
	\end{tabular}
\end{wraptable}

An external HiL reference device can be realized in many ways but should be qualified with proven methods to have consistent behavior.
Logic analyzers (\textbf{LA}) can be expensive and cannot inject signals or error cases.
USB-based bus controllers generally miss the timing requirements and features needed for testing.
Emulators and Simulation tools~(\textbf{ES}) are often limited to a certain platform and provide only limited features compared to the real hardware.
For example, Renode or QEMU emulators do not cover a large enough range of platforms and peripherals.
External slave devices (\textbf{SD}) such as sensors, which use a specific bus or peripheral, are quick to deploy for smoke tests but cannot exhaustively test the API or failure cases.
An \textbf{FPGA} solution would allow for extendable fine-grained control but can be more expensive, take more development time, and has limited off-the-shelf solutions.
An \textbf{MCU} solution sacrifices some control over the buses for ease of development and is available off-the-shelf.
\autoref{table:test_device_solutions} compares the relevant features of each solution indicating not supported, partially supported, or well supported.

Trying to keep up with fast IoT development cycles, agile processes such as Continuous Integration (CI), or applying checks on every change, are increasingly applied to embedded systems~\cite{m-cidal-19}.
While systems evolve, they are likely to incorporate new bugs and testing needs to follow the agile development.
Bugs may arrive with new features or new usage patterns that were not considered by the existing testing infrastructure.
In both cases, the test tool must allow for adding test capabilities quickly to cover those new features and also prevent future regressions.
To keep pace with rapid feature changes, we aim for a testing tool that is easy to acquire, adapt, and extend for the test tool users.
We see these objectives supported by off-the-shelf hardware with an open-source implementation.
Being able to look at the software implementation of the testing tool gives insight on how exactly tests are performed and what kind of constraints or implications are bound to it.
Developers possessing very detailed domain knowledge on devices or software implementations are provided with a clear path on how to transfer this knowledge into automated HiL tests.
Considering this, the MCU solution fits best given the constraints and domain knowledge of our target audience.

\section{System Overview of the HiL Environment}
\label{sec:sys}

Our proposed testing environment consists of an ensemble of three components and tooling (see \autoref{fig:Robot-FW-arch}).
All components and the associated infrastructure are designed for testing an embedded \ac{OS}.
Our implementations separate tester and testee  via an abstraction layer, which makes most implementations in our system agnostic to the \ac{OS} and devices~used.

The \tnode conducts the tests orchestrated by a common framework.
It executes the test firmware on the \ac{DUT}, and the external reference device, \philip.
\philip serves as a qualified reference to test peripheral API implementations across the supported \acp{DUT}.
All three devices interact through the wiring of peripherals and GPIOs.

To solve the challenges outlined in \autoref{sec:problem}, \philip allows for automated tests on various target boards, provides low maintenance and deployment costs.
\philip runs a single firmware that uses auto-generated code from a tool called \mmm.
The \mmm~processes a simple configuration file and easily extends the API of \philip allowing for the development of test capabilities to be agile.
We refer to \autoref{sec:philip} for details.

\subsection{Testing Abstraction Layer}
\label{sec:test_interface}

Coordination between \philip and the DUT happens without altering the firmware.
The peripheral APIs of the DUT are exposed through an interactive shell in a structured way.
This structure is shared with \philip allowing a coordinator to apply the same instructions to both \philip and the DUT.
Python wrappers simplify and unify the interface to the \tnode by providing classes with structured output enabling queries for statistics and benchmarking.

Coordination is not the only benefit of a structured testing language.
Following structured testing guidelines allows developers to write tests with a unified process for handling and executing tests.
Implementation of test logic independent of the firmware reduces the flash cycles that are needed for every test.
Exposing the API comes at a fixed overhead cost.
As the number of tests grows, the size of the firmware does not.
Our goal is to keep the constraints off the \ac{MCU} but leave them to the more capable \tnode, which can easily verify all advanced test options.

\subsection{The Structured Testing Interface}
\label{sec:structTI}
The structured interface is provided via the python wrappers.
This allows test API grammar to be defined and exploited.
Five conventions are applied to both \philip and the \ac{DUT} to simplify testing:

\begin{itemize}[noitemsep]
	\item[C1] The communication method is synchronous commands and responses.
	\item[C2] The response information is a dictionary adhering to a schema.
	\item[C3] Every response contains a result which will either be success, error, or timeout.
	\item[C4] Optional data returns simple, predefined types.
	\item[C5] Time-critical steps should be wrapped in a single synchronous command.
\end{itemize}

By following these basic conventions code reusability is increased.
Additionally, the test structure is unified and lowers the barrier for developers to understand existing tests and expand the test base.

\section{\philip: A Modifiable Reference Device}
\label{sec:philip}

\philip is a reference device for automated peripheral testing.
It consists of a nucleo-f103rb or bluepill board with open-source firmware.
\philip can use the MCU hardware to collect physical information from a \ac{DUT} similar to a logic analyzer but can also inject specific peripheral behaviors.
\philip uses a serial connection wrapped with \philippal to simplify interaction and provide an intelligent shell.

As illustrated by \autoref{fig:Robot-FW-arch}, a core component of \philip is the \mmm which feeds memory map information to \philip firmware, the \philippal interface, and documentation.
\philippal allows integration of a CI system and a developer to interact with the \philippal shell and read documentation from the memory map.

\subsection{\philip Objectives}
\label{sec:obj}
The goal of \philip is to have an extendable reproducible solution for testing real-world characteristics of the peripheral APIs of embedded devices (\ie UART, SPI, I2C, ADC, DAC, PWM, timers, GPIO).
The peripherals should be able to:
\begin{enumerate}[noitemsep]
	\item Read and write bytes and registers via I2C and SPI.
	\item Support different modes and addressing.
	\item Allow for various speeds (I2C 10--\SI{400}{\kilo\hertz}, SPI 0.1--\SI{5}{\mega\hertz}, UART 9600--\SI{115200}{\baud}).
	\item Support different register sizes (8 bit, 16 bit) with both big and little endianness.
	\item Track peripheral interactions such as bytes sent and received.
	\item Inject error signals and artificial delays.
	\item Estimate bus speeds for I2C, SPI with 5\% tolerance.
	\item Log timestamped GPIO events with a precision of \SI{200}{\nano\second}.
\end{enumerate}

The pinout on \philip is static so that rewiring is not needed when testing different peripherals.
The languages (C and Python) and tools used to develop \philip are familiar to developers testing with it.
\philip serves as a specific example of the general concepts for agile test tool development.

\paragraph{Qualification}
In an agile development environment, qualification must occur frequently and with little cost.
\philip enables this by taking an inexpensive piece of hardware, automating qualification with a single set of more costly or rented tools, and then distributing the inexpensive hardware to all test nodes.
This process is valuable in many situations, including
	 \one requiring many copies of reference hardware where purchasing off-the-shelf qualified equipment is too costly;
	 \two working with remote developers that require physical access to expensive qualified reference devices;
	 \three having occasional access to costly tools.

\subsection{\philip Firmware Implementation}
\label{sec:pfi}
\begin{figure}
	\centering
	\includegraphics[width=0.6\columnwidth]{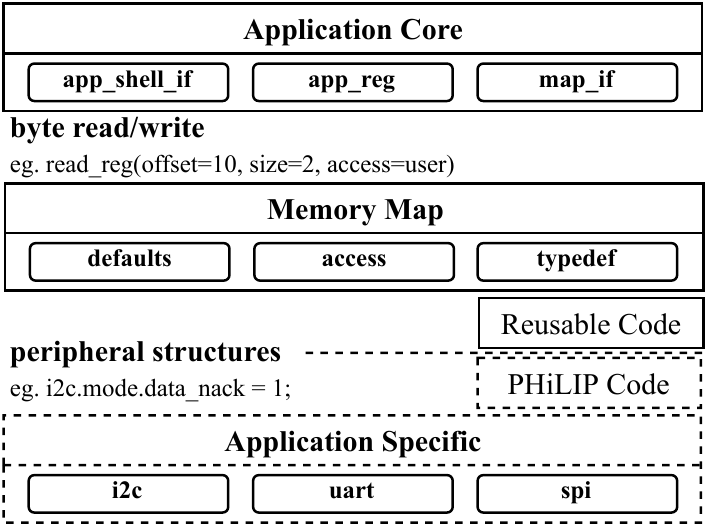}
	\caption{\philip firmware module interaction with array access from the application core and structure access from the peripheral modules}
	\label{fig:PHiLIP_firmware_arch}
\end{figure}

\philip firmware is designed to easily add peripheral functionality.
It separates the peripherals and application-specific functions from communication, parameter access logic, and the memory map as shown in \autoref{fig:PHiLIP_firmware_arch}.
The application core code and memory map definition process of \philip code is reusable in other projects and has versionable firmware components for structured communication.
The application-specific code in \philip implements peripheral instrumentation.
Without optimizing for size, the \philip application requires less than 36~kB and leaves at least 28~kB for future upgrades.

\paragraph{Application Core}
The application communication protocol provides a simple serial interface to read or write the parameters of the memory map as a byte array implemented in the firmware (see \autoref{table:protocol}).
The array is packed into a typedef structure allowing the C code to use descriptive names.
To support multiple simultaneous changes in configuration, the parameter access functions require that the execution command is called after all changes to the memory map are completed.
All parameter changes undergo access protection checks and safe handling (\eg disabling interrupts).
Accessing the memory map data as an array is valuable for peripherals that can read or write registers such as SPI and I2C, and then verify with a different interface.
\philip~contains 128 bytes of shared user data that can be accessed via both the peripherals and the \texttt{app\_shell\_if}.
The size and offsets of the memory map can change and the version command is used to identify the correct mapping.

\begin{table}
	\centering
	\caption{Basic \philip firmware protocol commands}
	\label{table:protocol}
	\begin{tabular}{lll}
		\toprule
		{\bf Command}              & {\bf Description}            & {\bf Example Return}\\
		\midrule
		\texttt{rr <index> <size>} & Read application registers   & \texttt{\{"data": 42, "result": 0\}}\\
		\texttt{wr <index> [data]} & Write application registers  & \texttt{\{"result": 0\}}\\
		\texttt{ex}                & Apply changes to registers   & \texttt{\{"result": 0\}}\\
		\texttt{-v}                & Print interface version      & \texttt{\{"version": "1.2.3", "result": 0\}}\\
		\bottomrule
	\end{tabular}
\end{table}

\paragraph{Time-critical Peripheral Event Handling}
The \philip firmware should allow all time-critical events to be stored for later access or prepared before the event occurs.
\philip uses peripheral hardware, interrupts, and polling to capture the information from events that occur.
Using the \ac{MCU} peripheral hardware allows a simpler implementation without requiring overclocking of \philip with respect to the \ac{DUT}.
Specific peripheral behaviors can be triggered by preparing a state for an expected \ac{DUT} action.
If several time-critical events occur before data can be accessed, then the information can be stored as counts, sums, or as an array of events.
For example, the GPIO module logs a timestamp per interrupt in a circular buffer that can be accessed after a series of rapid pin toggles.

\subsection{\philip Memory Map}
\label{sec:mmm}

To keep \philip easily adaptable while maintaining a low memory footprint, we developed the \mmm~
as a code generator for coordinating application information from a single configuration file.
This reduces human error when adding or changing runtime parameters, improves development speed, and the information can be fed into tests with various interfaces.
The JSON configuration file follows a schema that can provide named packed structures to embedded devices, and allows for documentation of the register map.
Structures and parameter properties such as type, array size, or testing flags are defined from the configuration file, as well as default values, access levels, and information for describing the parameters.
The registers are serialized and can be accessed as a structure (by name) or byte array (by address).

Describing the memory map based on parameter names with the respective types and sizes combines the versatility of named access with the efficiency of serialized packed memory.
The interface only needs to translate the name to an offset and size to get the information.
The simplicity of implementing only read and write register commands to deal with each parameter reduces bugs on the embedded device.

\begin{figure}
	\begin{tikzpicture}
		\node [write, label=above:\sffamily Module A] (ma_p1) {\sffamily write param 1};
		\node [write, right=4ex of ma_p1] (ma_pn) {\sffamily write param~N};
		\node [write, right=4ex of ma_pn] (ma_init) {\sffamily write init flag};
		\node [below=.5ex of ma_pn] (m_dots) {\vdots};
		\node [write, below=1ex of m_dots] (mz_pn) {\sffamily write param~M};
		\node [write, left=4ex of mz_pn, label=below:\sffamily Module Z] (mz_p1) {\sffamily write param 1};
		\node [write, right=4ex of mz_pn] (mz_init) {\sffamily write init flag};
		\node [execute, right=18ex of m_dots,yshift=-0.4ex] (ex) {\sffamily commit changes};
		\node [write, right=4ex of ex] (act) {\sffamily re-init modules};
		\path [line, dotted] (ma_p1) -- (ma_pn);
		\path [line] (ma_pn) -- (ma_init);
		\path [line] (ma_init) -| (ex);
		\path [line, dotted] (mz_p1) -- (mz_pn);
		\path [line] (mz_pn) -- (mz_init);
		\path [line] (mz_init) -| (ex);
		\path [line] (ex) -- (act);
	\end{tikzpicture}
	\caption{Setting parameters for modules (A--Z) on \philip.}
	\label{fig:commit}
\end{figure}
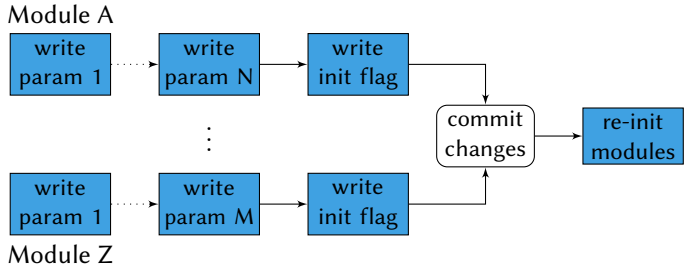

The generated output of the \mmm is C~style data and consumed by the firmware application.
By convention, parameters can be changed by writing registers, similar to \acp{MCU} or sensors.
To initiate a change in properties, for example, altering the I2C address on \philip, an initialization bit should be set before calling the command to execute changes.
This allows for a peripheral to be configured only once, preventing possible initialization sequence errors.
\autoref{fig:commit} shows an example of changing many parameters and executing the changes.

\subsection{The Abstraction \philippal}
\label{sec:abstphilippal}
\philippal provides a Python wrapper that implements the structured testing interface outlined in \autoref{sec:structTI}.
It takes the basic firmware protocol commands and maps bytes back to structure members using a CSV file generated from the \mmm.
\philippal first checks the version then correlates that version to a specific mapping.
As a result, the memory map can easily change or add parameters while maintaining backward compatibility for named access.
\philippal also provides the documentation of functions, validation of parameters, default arguments, and parsing of values, keeping this out of \philip firmware.
There are over 270 named fields that can be written or read in the map that corresponds to settings for parameters implemented in the \philip~firmware.
For example, the \texttt{i2c.r\_count} contains the number of bytes read from the I2C register.
The logic is implemented in the firmware, counting the number of I2C data ready hardware interrupts that occur and storing it in the mapped C structure.

\begin{table*}
	\centering
	\caption{Example showing how data traverses \philip abstraction layers via name-mapped parameter access}
	\label{table:protocolphil}
 \begin{adjustbox}{max width=\textwidth}
	\begin{tabular}{>{\bfseries}lll}
		\toprule
		Layer & \textbf{Request} & \textbf{Response}\\
		\midrule
		\philippal &
		\multicolumn{1}{r}{\texttt{phil.read\_reg("i2c.r\_count")}} \vline &
		\makecell[l]{\texttt{\{"cmd": ["read\_reg(i2c.r\_count,0,1)"],}
			\texttt{"data": [1], "result": "Success"\}}} \\
		Serial port &
		\multicolumn{1}{r}{\texttt{rr 334 1}} \vline
		&
		\multicolumn{1}{l}{\texttt{\{"data": 1,"result": 0\}}} \\
		\philip FW
		& \multicolumn{2}{l}{
			\texttt{printf("\{\textbackslash"data\textbackslash":\%u,\textbackslash"result\textbackslash":0\}\textbackslash n", read\_regs(334,1));\hspace{4cm}}
		} \\
		\bottomrule
	\end{tabular}
\end{adjustbox}
\end{table*}

Along with a Python class, \philippal provides a shell to assist developers in manual debugging with features such as autocompletion, self-documentation, and helper functions.
\autoref{table:protocolphil} shows name-based parameter access via \philippal being converted to addresses and offsets.
\philippal looks up the offset and sizes from the named parameter based on the versioned memory map indicated by \philip, then writes the command via the serial port to \philip.
\philip in turn, either prepares, applies, or reports the parameters.
In this example, reporting the requested value \texttt{i2c.r\_count} via the serial port, which then gets parsed by \philippal according to the datatype indicated by the map.
The \texttt{i2c.r\_count} parameter in \philip gets updated via I2C reads from the \ac{DUT} and are only fetched from \philip by the host computer when needed with a \texttt{read\_reg("i2c.r\_count")} command.

\subsection{Adding New Testing Capabilities to \philip}

The agile process to adopt a new test capability takes five steps:
\begin{enumerate}[noitemsep]
	\item Identify the parameter(s) needed based on a bug or issue.
	\item Add the parameter into the \mmm configuration file and generate a new map.
	\item Implement functionality on the given module.
	\item Qualify the parameter on \philip.
	\item Release the new firmware and Python package.
\end{enumerate}

If \philip cannot provide a way to either measure or induce the state where an issue occurred then a parameter is added to the memory map configuration file, \eg forcing a data NACK on the I2C bus.
The parameter has a descriptive name and other valuable information such as access level or unique flags.
A new memory map is generated that validates the JSON syntax and schema, recalculating the sizes and offsets of each parameter.
C code and a CSV file are exported to the sources.
Functionality is implemented in the C code for that given module, \eg if the \texttt{i2c.nack\_data} parameter is enabled \philip should set the \texttt{I2C\_CR1\_ACK} bit to 0.
An automated qualification procedure based on standard tools then validates the expected behavior.
Thereafter, the firmware can be released along with a Python package containing the new memory map data.

\section{\philip Performing Multi-platform HiL Testing}
\label{sec:tests}

Automated and platform-independent tests are created using \philip for the \ac{DUT} (see \autoref{sec:sys}).
The design of the testing environment takes advantage of the structured testing interface and existing tools where available, adding custom implementations where needed.
The tests can be run by developers or integrated with a CI system.

\begin{table}
\centering
	\caption{Comparison of build sizes on nucleo-f103rb board}
	\begin{tabular}{>{\bfseries}lcc}
		\toprule
		{\bf Test} & {\bf ROM [Bytes]} & {\bf RAM [Bytes]} \\
		\midrule
		periph\_i2c		& 19,780 & 2,656 \\
		periph\_gpio	& 13,828 & 2,520 \\
		periph\_spi		& 20,268 & 2,816 \\
		periph\_uart	& 20,672 & 4,536 \\
		periph\_timer	& 16,624 & 2,548 \\
		\bottomrule
	\end{tabular}
	\label{table:test_build_sizes}
\end{table}

\subsection{Testing RIOT OS Peripherals}

The \ac{DUT} firmware is implemented in C on RIOT~OS, which enables the multi-platform testing based on the \ac{DUT} \ac{HAL}.
Writing tests in this way can save flash cycles and limit code size as more tests are added.
\autoref{table:test_build_sizes} shows build sizes of the peripheral-based tests.
We group all tests that share similar properties in three groups, infrastructure testing, bus testing, and timer testing.

\paragraph{Infrastructure Testing}
Errors can occur from infrastructure components or setup.
Tests ensure that the infrastructure is operating properly within the testing system.
There must be a connection to the \ac{DUT} and \philip, opening a connection and sending a sync message can resolve this.
The wiring must be correct, thus toggling the GPIO of each wire can be used to verify the wiring.
The flashing tool for the \ac{DUT} must be functioning, reading a descriptor of the firmware can be a way to check the correct firmware is flashed on the \ac{DUT}.

\paragraph{Bus Testing}
\label{sec:bustesting}
A peripheral bus is stateful and involves exchanges between devices.
Since hardware registers need to be set and cleared by external interactions to introduce persistent states or race conditions, simply completing code coverage is not exhaustive.
\philip can be configured to inject and check for errors related to the time and state of SPI, UART, and I2C buses.
For example, it can alter clock stretching, flip bits, or record total interaction time.
When the \ac{DUT} executes an action to be tested, \philip interacts with the DUT according to the configuration and collects metadata on the interaction.
This information can be queried later from \philip via the host.
We consider the following five basic~tests:
\smallskip

\textit{Initialization Tests:} Initializing and acquiring any bus lock including powering on the hardware.

\textit{Usage Tests:} Read or write operations using the default configuration.

\textit{Mode Tests:} Varying the modes and settings of the bus and ensuring interactions are correct.

\textit{Negative Tests:} Check improper configurations return appropriate error messages.

\textit{Recovery Tests:} Check bus recovery after forcing an error state.

\smallskip
\begin{minipage}{0.97\columnwidth}
\begin{lstlisting}[style=PyStyle,
	caption={Example testing indication of I2C NACK condition},
	label={lst:i2cnack},
	columns=fullflexible]
philip.write_and_execute("i2c.mode.nack_data", 1)
response = dut.i2c_read_reg(PHILIP_ADDR, 0)
assert response["result"] == ERROR_RESULT
assert response["data"] == -EIO
\end{lstlisting}
\end{minipage}
\smallskip

\noindent \autoref{lst:i2cnack} shows an example of \philip injecting a challenging behavior for a test.
First \philip is prepared to NACK only data bytes in an I2C frame.
Then the \ac{DUT} executes an I2C read register command on \philip.
After the I2C is finished, the \ac{DUT} returns the result to the \tnode.
The expected result is an error with the \texttt{-EIO} error code indicating a NACK on the data has occurred.

\begin{minipage}{0.97\columnwidth}
\begin{lstlisting}[style=PyStyle,
	caption={Example for asserting metadata of I2C operation},
	label={lst:i2cprop},
	columns=fullflexible]
	response = dut.i2c_read_reg(PHILIP_ADDR, 0)
	assert response["result"] == SUCCESS_RESULT
	assert response["data"] == philip.read_reg("user_reg", 0)["data"]
	assert philip.read_reg("i2c.r_count")["data"] == 1
	assert philip.read_reg("i2c.w_count")["data"] == 1
\end{lstlisting}
\end{minipage}
\smallskip

\autoref{lst:i2cprop} collects metadata of an I2C read register command from \philip.
First, the \ac{DUT} executes an I2C read register command on \philip and stores the response.
The result should be successful and the data of that register should match data from \philip.
Additional properties of the I2C read register command are verified such as the number of bytes read from and written to \philip.

Initialization tests can catch bugs with powering and configuring the peripheral clock which may prevent startup.
Usage tests verify the correctness of basic operation in corner case conditions like maximum transfer size.
Mode tests expose unimplemented or wrongly implemented configuration options.
Negative tests probe if a module handles unexpected conditions appropriately.
With recovery tests it is possible to find bugs that occur only after rare fault conditions.

\paragraph{Timer Testing}
Timer tests are challenging, truly concurrent operations.
Events are generated at very high rates and require precise timing.
Isolated testing solely on the \ac{DUT} is infeasible as a reliable internal reference time is often missing.
Complex code for measuring and analyzing corrections on the \ac{DUT} induces side effects, which cannot reliably be quantified nor compensated across platforms.
Using pure command-response communication between \tnode and \ac{DUT} does not allow for precise timing either, because the communication channel (\eg  UART) and command parsing induce significant delays and jitter.
Therefore, a clear separation between test overheads and time-critical hardware instrumentation is needed.

With \philip, this is achieved by asynchronously logging GPIO signal events with timestamps.
DUT timer operations are instrumented for signaling via designated pins.
The pins are dynamically configured on \philip before the test execution.
Thereafter, timestamped GPIO-traces are acquired via \philippal, and measurements such as frequency, drift, jitter, and accuracy are calculated and compared to tolerance values.
Respective limits on timing accuracy of the current \philip implementation are detailed in \autoref{sect:phil:timing}.

\begin{figure*}
\begin{subfigure}{0.48\columnwidth}
  \centering
  \includegraphics[]{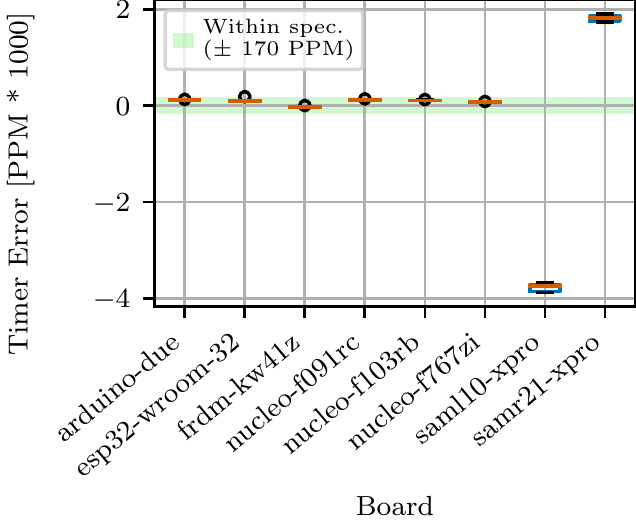}
  \caption{Timer errors compared to oscillator specification threshold across multiple different boards}
  \label{fig:timerppm}
  \end{subfigure}
\hfill
\begin{subfigure}{0.48\columnwidth}
  \centering
  \includegraphics[]{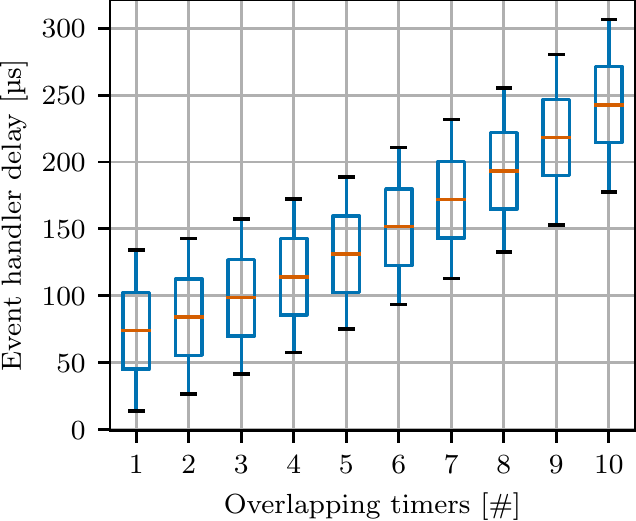}
  \caption{Delays of timer event handler on nucleo-f091rc when multiple timer targets overlap}
  \label{fig:timerhandlerdelay}
\end{subfigure}
\caption{Examples of two timer tests, accuracy, and delay}
\vspace{-3mm}
\end{figure*}

\autoref{fig:timerppm} shows sample results of a cross-platform test on timer accuracy.
The threshold given as \SI{170}{PPM} (Parts Per Million) combines datasheet limits for crystal oscillator accuracy, stability, and aging values of the \ac{DUT} and \philip.
Two of the tested boards clearly do not comply with this specification.
Likely sources for such errors are incorrectly configured clock sources (\eg using an internal resistor-capacitor oscillator instead of a crystal), improper trimming configuration of the crystal oscillator, or faulty prescaler configurations.

A further example grounded on precise microsecond-scale measurements is the assessment of worst-case delay limits under specific edge cases, \eg \texttt{n} virtual timers triggering at the same time.
\autoref{fig:timerhandlerdelay} exhibits the upper bound for the timer scaling behavior of virtual (software multiplexed) timers on nucleo-f091rc board.
The plot shows a linear scaling behavior, indicating a maximum delay close to \SI{300}{\micro\second} when ten timers are scheduled for the same target time.

\subsection{Developer Testing Locally}
\label{sec:devtest}
The automated test suites can be run by developers locally if the boards or wiring are not supported in the CI.
The setup requires \philip firmware to be flashed on a nucleo-f103rb or bluepill board.
The testing repository along with python requirements must be installed.
Only the wiring needed for the specific test must be connected from the \ac{DUT} to \philip.
If wiring differs from the CI boards then it must be input to the test environment.
A single \texttt{make} command allows the tests to be flashed and executed.

\subsection{Flexible CI Integration}
\label{sec:citest}
\begin{figure}
	\centering
	\includegraphics[width=0.5\columnwidth,trim=0 0 0 0,clip]{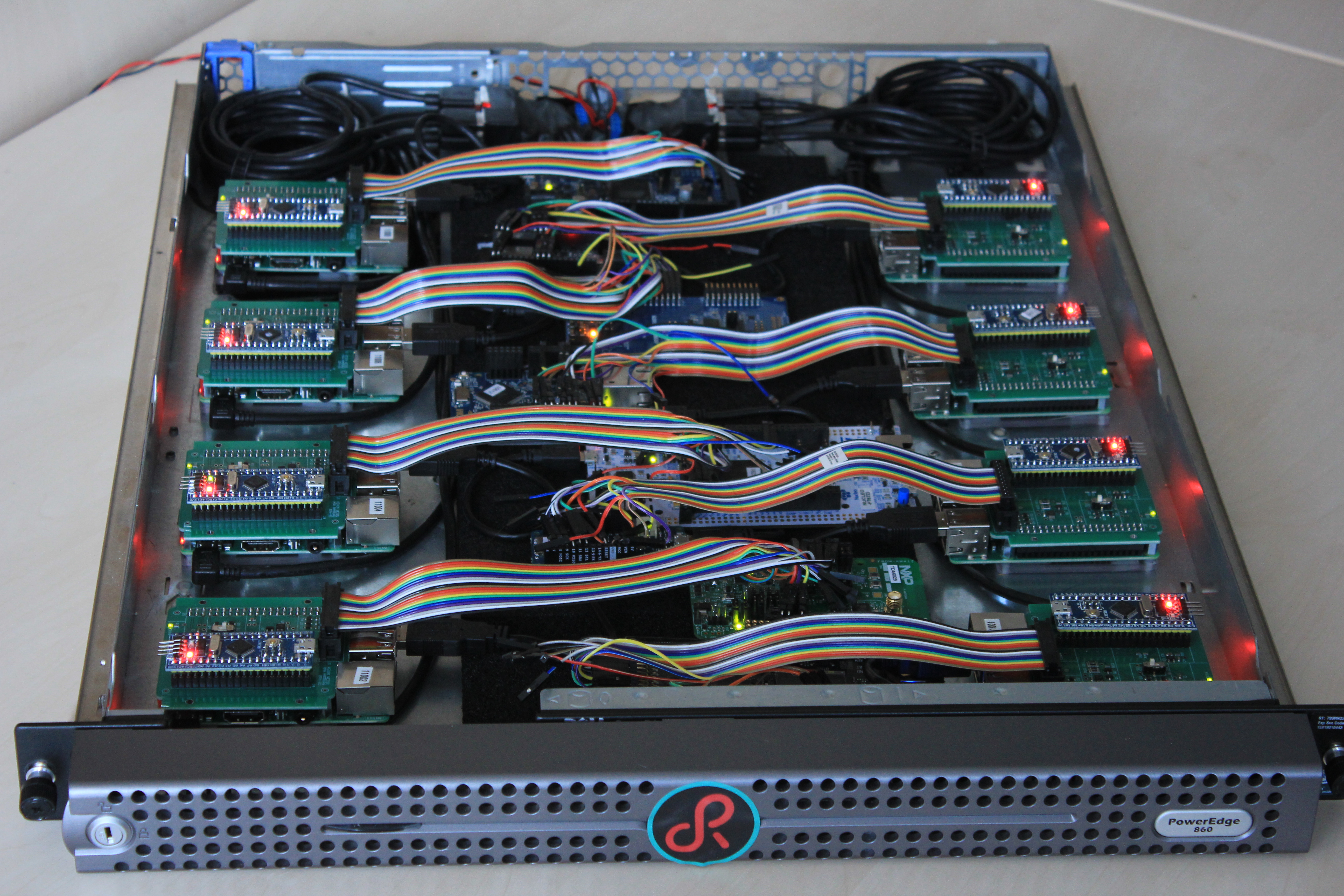}
	\caption{Up to 8~test nodes can be packed in 19" server rack}
	\label{fig:rack_front}
\end{figure}

\philip can be used from off-the-shelf components, however, a custom board was created to ease CI deployment.
A CI \tnode consists of the custom board that provides connections between a Raspberry Pi and \philip and a standard 20 pin ribbon connector to the \ac{DUT} (see \autoref{fig:rack_front}).
This allows for the simple wiring of \acp{DUT} in different form factors without developing specific breakout boards.
A test is provided to ensure wires are correctly routed.
The custom board also provides basic power measurement and control tools to help with low power testing, some protection circuitry, and signal conditioning.

Each \tnode is responsible for testing a single \ac{DUT} in the CI setup.
The 1:1 ratio of \tnode to \ac{DUT} was chosen over a 1:n ratio for multiple reasons.
Managing the \tnode environment is simplified.
Downtimes of a single \tnode become less critical.
DUTs do not need to share USB bandwidth.
Computationally heavy tasks on the \tnode will not affect other running tests.
Building is done on separate servers with docker and does not require any special hardware.

In our current setup, tests run on 22 different boards\footnote{up-to-date board list:
\url{https://ci.riot-os.org/hil/labelsdashboard/}
}
of various form factors, vendors, and CPU architectures.
We test nine CPU architectures (AVR, cortex-m0+, cortex-m0, cortex-m3, cortex-m4, cortex-m7, cortex-m23, esp32, esp8266) three peripheral buses (I2C, SPI, UART), timers, and GPIO.

At the software side, our CI environment is based on three common open-source tools.
\one Jenkins steers continuous integration across multiple test nodes.
\two Robot Framework\footnote{\url{https://robotframework.org}} implements the test logic with its keyword-based syntax that can easily be extended and adapted to a specific domain.
\three Ansible is used to add and manage each \tnode.
Text execution is orchestrated by Jenkins allowing tests to be triggered manually, every night, and on every change of the test code.

\section{Evaluation}
\label{sec:evaluation}

\philip has powered continuous HiL testing in RIOT for over a year with over 200 stable testing cycles.
1519 tests are executed per night for 22~unique boards, taking less than 45 minutes.
In comparison, a developer may need the equivalent time to manually set up and execute a test on a single board.

In the following, we evaluate \philip in detail from five perspectives.
\one A case study of using \philip during a large I2C rework;
\two the timing constraints of measuring  and executing tests;
\three system overhead introduced by  tools, the \mmm, and \philippal;
\four the memory consumption exposing an API vs. hardcoding test cases;
\five  The costs of the CI infrastructure and developer usage.

\subsection{Impact: The I2C Rework Use Case}
\label{sec:itiruc}
%
%
%
%
%
%

\philip was used  during a two-month rework of the I2C peripheral in RIOT.
A small python script\footnote{
\url{https://github.com/RIOT-OS/RIOT/pull/9409}
} was used to run automatic tests on developer machines.
This required wiring of the I2C pins for each \ac{DUT} to \philip.
The script initialized both \philip and the \ac{DUT}, then ran a number of checks with different parameters such as varying addresses, flags, sizes, \etc to check the new API being introduced.
It discovered and prevented the bugs shown in \autoref{table:bugs_found}.
Bugs with high priority typically mean they exist in master or are about to be merged into master.
Bugs with higher severity affect the current tests and drivers whereas minor severity bugs are edge cases that may occur in external applications.

\begin{table}
	\centering
	\caption{List of bugs discovered by \philip during an I2C rework. CWE (Common Weakness Enumeration) provided by \url{https://www.mitre.org/}}
	\label{table:bugs_found}
	\begin{tabular}{>{\bfseries}llllll}
		\toprule
		{\bf Family} & {\bf Boards} & {\bf Priority} & {\bf Severity} & {\bf CWE} & {\bf Description} \\
		\midrule
		sam0 & 5 & medium 	& moderate & 474 & 	Inconsistent Function Implementation \\
		\midrule
		sam0 & 5 & medium 	& moderate & 394 & Unexpected Status Code or Return Value	\\
		\midrule
		atmega & 3 & low 	&  major & 480 & Use of Incorrect Operator\\
		\midrule
		cc2538 & 5 & medium		&  major & 460 & Improper Cleanup on Thrown Exception\\
		\midrule
		stm32 & 18 & low & minor & 394 & Unexpected Status Code or Return Value\\
		\midrule
		stm32 & 18 & high & major & 835 & Loop with Unreachable Exit Condition\\
		\bottomrule
	\end{tabular}
\end{table}

The hidden CWE474 bug was discovered on the \texttt{sam0} platform.
It was a physical read of an extra byte, leading to additional time used on the I2C line as well as potentially increasing a register pointer on the secondary device.
Since the byte was discarded in software, other tests with sensors were not able to detect the failure.
The \texttt{i2c.r\_count} parameter provided by \philip about how many bytes were read from the I2C bus uncovered the bug for the whole sam0 platform.
The test in \autoref{lst:i2cprop} discovered that the read byte count was 2 instead of 1.
A CWE394 bug was also discovered in the process.
In this case, the return value was always successful, even when failing, showing the importance of negative tests.

Testing on the \texttt{atmega} platform, a CWE480 bug caused I2C register writing to fail.
This was due to missing an inversion of the bitfield when checking a status bit causing incorrect state transitions.
The CWE460 discovered on the \texttt{cc2538} platform caused lockup after reading from a missing address.
This was due to the internal hardware not clearing its error state before issuing another command.
Both CWE394 and CEW835 were discovered on the \texttt{stm32} platform prevented multiple I2C register writes.
This was due to stop signals being sent when busy causing the state to hang when attempting to issue another write command.

\subsection{Timing Constraints}

\paragraph{\philip Qualified Timing Constraints}
\label{sect:phil:timing}
Temporal accuracy is a relevant constraint for \philip as the timer tests described in \autoref{sec:tests} for instance, need high-resolution measurement methods.
\philip's capabilities are evaluated against accurate measurement equipment as part of the qualification procedure outlined in \autoref{sec:obj}.
This is done by setting the measurement equipment to toggle a pin at different rates and verifying the readings of \philip.
The limits of different time capture methods supported by \philip are summarized in \autoref{table:timingqual}.
The minimum time between two consecutive logging events is denoted by $t_{min}$.
The maximum accepted jitter of the time measurements is listed as $t_{jitter}$.
DMA instrumented timer capturing performs best but, due to \ac{MCU} hardware limitations, can only capture either rising or falling edges.
The Timer \ac{IRQ} variant can be triggered by rising and falling edges but relies on slower CPU instructions to read timer values.
Both variants using designated timer hardware allow high precision but restrict the number of events by the associated buffer size to 128.
The GPIO \ac{IRQ} sampling uses interrupts of GPIO hardware to log timestamps directly to the memory map.
With both edges and virtually infinite sampling duration, this gives the highest flexibility but limits precision.

\begin{table}
	\centering
	\caption{Comparison of timing constraints for different instrumentation methods provided by \philip.}
	\label{table:timingqual}
	\begin{tabular}{llll}
		\toprule
		\textbf{Measurement Method} & \vrule & \textbf{$t_{min}$} & \textbf{$t_{jitter}$}\\
		\toprule
		Timer Capture DMA           & \vrule & 200 ns             & 28 ns  \\
		Timer Capture IRQ           & \vrule & 1 us               & 200 ns \\
		GPIO IRQ sampling           & \vrule & 10 us              & 600 ns \\
		\bottomrule
	\end{tabular}
\end{table}

\paragraph{Command Timing Constraints}
\philip solves the need for strict timing requirements between the \tnode and \ac{DUT} by measuring time-critical parameters locally and reporting them later to the \tnode.
The completion time varies depending on the command and communication method.

\begin{figure}
\begin{minipage}{0.48\columnwidth}
\includegraphics[]{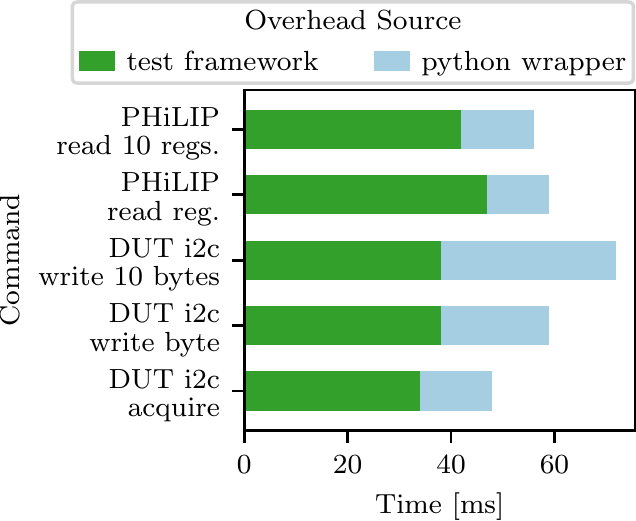}
\caption{Timing of framework overhead and command time for a nucleo-f103rb using the periph i2c test.}
\label{fig:commandbreakdown}
\end{minipage}
\hfill
\begin{minipage}{0.48\columnwidth}
\includegraphics[]{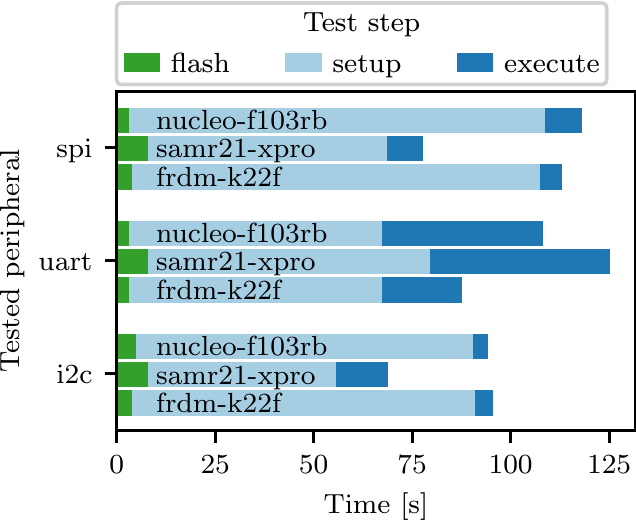}
\caption{Average test step time per board for a given tests.}
\label{fig:timebreakdown}
\end{minipage}
\end{figure}

\autoref{fig:commandbreakdown} shows the execution time of various commands collected from Robot Framework test artifacts produced in the nightly CI runs.
The time for the python instruction includes sending the command from the \tnode to the \ac{DUT}, its execution, returning results to the \tnode, and finally parsing it.
The framework overhead is the time Robot Framework takes to log the steps and check the results.
This depends on the  speed of the \tnode, \ie a relatively slow Raspberry Pi 3.

These times help determine the limits using synchronous commands and  when time-critical events should be offloaded to a grouped command.
With this setup, anything that requires timing below milliseconds should be offloaded.

\subsection{Test System Overhead}
\label{sec:tso}

\paragraph{Duration of Tests}
We present results for three test suites on three different boards in \autoref{fig:timebreakdown} selected to show the largest variations of time.
Each value is averaged over 30~nightly CI runs though time variances between runs are negligible.
We focus on the time the microcontroller is in use excluding timing metrics related to the \tnode.
The setup and execution time are captured from the Robot Framework output whereas the flash time is taken from the CI logs.
The flash time depends on the binary size and the speed of the flasher.
Some boards have slow flashers due to low baud rate UART communication.
Flash times range between 3 and 10~s.

The test setup time is the longest across all cases.
A setup phase occurs before each test and resets both the \ac{DUT} and \philip, then establishes a connection by reading the expected firmware version.
The reset times of a \ac{DUT} vary due to board-specific delays introduced by a failed initial connection attempt due to the spurious bits of the UART peripheral being reset.
Depending on the bootloader, a silent period may be needed before connecting, for example, the Arduino bootloader requires 2.6 seconds after resetting.

During the test execution time, the \ac{DUT} and \philip are interacting via peripherals.
This includes the time to send a command from the \tnode to the \ac{DUT}, to execute the API call on the \ac{DUT}, to return the result  to the \tnode, and parse that result.
The UART tests show a longer execution time due to the \ac{DUT} sending a large amount of data at the baud rate limit. The frdm-k22f board finishes faster as it skips tests of unsupported modes.

\paragraph{Memory Map Overhead}
The cost of using the \mmm can be shown by evaluating both the memory footprint and the speed at which the data can be accessed.
\autoref{table:mm_overhead} shows the differences in the overhead of \philip firmware between address-based access using the size and offset versus name-based access using a name and decoding the information in the firmware.
\philip's memory map contains 273 parameters taking a minimum of 1841 bytes, a total of 2048 bytes with padding.

\begin{table}
	\centering
	\caption{Comparison of memory and parse time for address-based parameter access vs. named-based parameter access.}
	\label{table:mm_overhead}
	\begin{tabular}{>{\bfseries}lll}
		\toprule
		Access & {\bf Flash (kB)} & {\bf Parse Time (\SI{}{\micro\second})} \\
		\midrule
		By Address & 31.4 & 22.4 	\\
		\midrule
		By Name & 40.8 & 38.6 \\
		\bottomrule
	\end{tabular}
\end{table}

We compare a variant of \philip firmware that allows the memory map to be stored inside the firmware instead of using \philippal to decode the map.
Reading and writing parameters by name rather than address-based reads with size and offset increases flash size by 9352 bytes.
The ability to access the map properties through firmware adds 384 bytes but allows \philippal to use the map without previous information.
Accessing with names also adds to the response time due to increased parsing complexity.
\philip is instrumented to toggle a GPIO pin when a command is received resulting in \SI{22.4}{\micro\second} to parse a \texttt{rr 0 1} command mentioned in \autoref{table:protocol}, that reads one byte from the user register.
Reading with a name, for example, \texttt{r user\_reg 0}, results in \SI{38.55}{\micro\second} parsing time.

\subsection{Memory Usage}
\newcommand{\verbint}{\textit{verbose interactive}\xspace}
\newcommand{\minint}{\textit{minimal interactive}\xspace}
\newcommand{\nonint}{\textit{verbose output only}\xspace}

The memory footprint is evaluated by comparing build sizes for three variants of the \ac{DUT} interface, differing in interaction functionality and verbosity.
The two interactive variants communicate synchronously between the \tnode and the \ac{DUT}, where the first one uses a text-based, human-readable shell and the second uses more concise binary encoding.
We refer to both cases as \verbint and \minint respectively.
The third, self-contained variant still outputs verbosely  but does not require input from the \tnode.
Therefore, we refer to it by \nonint in the following.

\begin{figure}[h]
	\centering
	\includegraphics{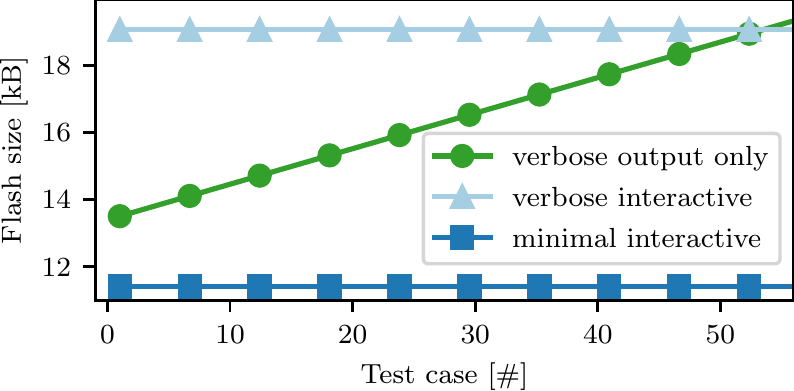}
	\caption{Flash memory size for DUT test firmware vs. the number of test cases.}
	\label{fig:memconsump}
\end{figure}

\autoref{fig:memconsump} relates the memory usage of all three approaches based on an I2C test firmware for the nucleo-f103rb considering an empirical average memory size increment of around 106 bytes per test case.
The reduced memory usage of the \minint case always has an advantage over the more verbose alternatives.
But even the \verbint version becomes more memory efficient than the self-contained firmware after surpassing the break-even point around 53 test cases.
This shows that, despite its additional code to expose \ac{DUT} functionality, offloading test cases to the \tnode saves memory for a large number of test cases.

\subsection{Cost of Testing on Hardware}
\label{sec:coth}
\newcommand{\eurosym}{\textup{\euro}}

We analyze the cost associated with testing on real hardware in \autoref{table:cc_breakdown} in terms of capital expenditure (CAPEX) and operational expenditure (OPEX) by considering the main expenses of two usage scenarios for \philip: the minimal developer desktop setup overhead as shown in \autoref{sec:devtest} and the overhead from automated testing via dedicated CI infrastructure as shown in \autoref{sec:citest}.
Common to both scenarios are the costs related to the \ac{DUT} which are displayed separately.

\paragraph{DUT}
The average cost of our deployed \acp{DUT} is 40 \eurosym, with the most inexpensive board being the esp8266-esp-12x at 7 \eurosym, and the most expensive board being the frdm-k64f 136 \eurosym.
The maintenance interval of the \ac{DUT} is determined by its flash endurance.
To obtain a realistic estimate on \ac{DUT} lifetime, we survey manufacturer datasheets of 7 unique boards supported by the RIOT CI
(arduino-due, arduino-mega2560, frdm-k22f, nucleo-f103rb, remote-revb, samr21-xpro, slstk3401a).
The worst-case and most common flash cycle range is 10k.
The OPEX of the \ac{DUT} consists of the replacement cost that occurs after 1250 full test runs consisting of 8 flashes per run.

\paragraph{Testing on the Desk}
Based on empirical values, students working with RIOT OS and \philip need approximately 2 hours to initially set up the hardware and testing environment which is reduced to 30 minutes once they are familiar with the process.
The operational expenditure for running the full test suite with the minimal setup is heavily dominated by only this labor time, effectively marginalizing other operational costs in this scenario.
The capital expenditure overhead is the cost of a nucleo-f103rb kit, some wiring, and the time to flash \philip firmware on it.

\paragraph{CI Testing}
The operational expenditure per CI run consists of build server time, around 2 minutes per run\footnote{Based on 4 CPU, \SI{16}{\giga\byte} memory, On-demand from \url{https://calculator.aws/\#/createCalculator/EC2}}, and labor costs for maintenance of replacing the \ac{DUT}, which take around an hour per maintenance interval.
Compared to the cost of the \acp{DUT}, the initial \philip and \tnode investment fits in the cost range of the \acp{DUT}.
The operational costs of the CI also are similar to the cost of replacing the \acp{DUT} at the maintenance interval.

\paragraph{Cost Reduction Potential}
The flash cycle limitation could be overcome by executing tests from RAM at the cost of accurately representing target devices in production.
Execution from RAM would additionally require per-target customization of linker scripts, limit the allowable code size, and alter timing due to missing flash wait states.
The CAPEX of the CI could be reduced by using more \acp{DUT} per \tnode, however, it would only pay off with a large number of \acp{DUT} due to the additional complexity implying further maintenance and development costs.

\begin{table}[h]
	\centering
	\caption{Cost breakdown of HiL usage for desktop setup overhead, CI overhead, and base \ac{DUT} cost.}
	\label{table:cc_breakdown}
	\begin{tabular}{>{\bfseries}lccc}
		\toprule
		& \textbf{Desktop}  & \textbf{CI} & \textbf{DUT} \\
		\midrule
		\textbf{OPEX}  & 30 [$\frac{mins}{run}$]  & 0.05 [$\frac{\eurosym}{run}$] & 0.01 to 0.12 [$\frac{\eurosym}{run}$]\\
		\midrule
		\textbf{CAPEX} & 10 [\eurosym] & 80 [\eurosym] & 7 to 136 [\eurosym]\\
		\bottomrule
	\end{tabular}
\end{table}







\section{Related Work}
\label{sec:related_work}

Testing with HiL fills important gaps as discussed in \autoref{sec:problem}.

\paragraph{On-target Testing}
Strandberg~\cite{s-aslst-18} points at significant time barriers due to the duration of complex tests.
Tight hardware coupling of the tested software further limits the availability of the environment  as it requires direct hardware access.
Orthogonal to our solution, Strandberg focuses on a management layer for optimized test selection and allocation of multiple networked devices.
We target testing \acp{HAL} with physical interactions, therefore, focusing on test execution with generic hardware instrumentation.

M{\r a}rtensson~\cite{m-cidal-19} considers test execution on real hardware as a hurdle for CI/CD integration because of limited access to custom target devices.
Even though simulation-based testing can take over selected test tasks, \eg checking functional stability~\cite{gg-ssim-19}, it is not suited to test whether the software will correctly operate on hardware.
Our solution overcomes these problems with on-target testing and shared access through CI, nightly, and on-demand testing workflow integration.
Instead of verifying  a specific application on a single device, we perform multi-platform testing on a wide range of heterogeneous \acp{DUT} to guarantee platform-independent firmware has deterministic behavior on all target devices.

\paragraph{Tools for Testing Embedded IoT Systems}
Specialized solutions cover domains from the custom silicon layer~\cite{cadence} and printed circuit board schematics~\cite{proteus} to system modeling, simulation, and automatic code generation~\cite{simulink}.
Common tools for testing usability, reliability, and compatibility of IoT devices, however, do not provide a solution for automated testing of tightly hardware-coupled software~\cite{mbqa-stti-18}.
\emph{IoT-TaaS}~\cite{kahbl-itpit-18} deals with coordination of interoperability and conformance testing of higher layers (\ie network protocols).
Testbeds are a common approach to test embedded software on real hardware under realistic conditions and in larger setups~\cite{hkww-tsrtwiesn-06,lmdbt-tfttsbwsn-15,abfhm-filso-15, gcz-sptbi-19}.
While they are mostly used for experimentation and performance evaluations of protocols and applications, some also explicitly focus on the integration of heterogeneous multi-node setups~\cite{lfzws-ftdst-13, tdsmb-fmmtv-20}.
Aiming for improved reliability, Woehrle~\etal~\cite{wpbt-irwsn-07} propose a distributed testing framework that combines simulation and testbed support tailored to the development process of wireless sensor networks.
Tools such as \textit{Greentea} and \textit{Icetea}~\cite{mbed-testing}, \textit{LAVA}~\cite{lava}, and \textit{ICAT}~\cite{cllt-iidct-18} focus on deployment, execution, or operative management abstractions, leaving measurement of physical hardware interactions limited or out-of-scope, or target mobiles instead of constrained IoT devices.

Izinto~\cite{plf-ipitf-18} is a pattern-based test automation framework for integration testing.
This solution does not require technical knowledge about the tested system but is limited to user perception use cases, whereas our work targets low-level misbehavior.

\paragraph{The Importance of Testing Peripheral Abstractions in Embedded Software}
System standards such as POSIX are not applicable for deriving test items for hardware interfaces~\cite{scs-itmhs-07}.
Therefore, Sung~\etal~\cite{scs-itmhs-07} contribute a test model, defining a list of test features for interfaces of \ac{OS} and hardware layers.
Seo~\etal~\cite{skcl-wssit-08} used this model to show that the likelihood of finding errors is significantly higher in interface functions that cross heterogeneous abstraction layers.
Their analysis further indicates that bugs in this type of code are harder to discover with classical unit testing approaches.
Justitia~\cite{ssck-aeste-07} was developed using the same model to automate the identification of interfaces to be tested together with generating and executing test cases.
The employed fault detection method is tailored very closely to a specific target platform and only applies to time-invariant errors (\eg in memory management and allocation).
It poses significant limits when timing critical operations with true hardware concurrency and connections to the physical world are considered.
Although the \ac{DUT} instrumentation principle is orthogonal, our focus on peripheral abstractions explicitly targets the important areas where execution flows cross layer boundaries.

Feng~\etal~\cite{fml-pshft-20} automate implementations of peripheral models that approximate the behavior of real peripherals for use in emulation-based fuzzy testing.
In contrast to our work, their approach aims for generic testing of the firmware on top of peripherals and is therefore neither able to model arbitrary hardware, DMA peripherals in particular, nor correct physical layer interactions.

\paragraph{Testing Physical Interactions With HiL}
HiL solutions are commonly used in the automotive domain~\cite{tessl-gfky-18,kr-mbtes-12, vgklm-hlsap-14,dspace}.
Vejlupek~\etal~\cite{vgklm-hlsap-14} propose a hybrid HiL approach, \ie stimuli signal injection allows them to use a parameterized model instead of a complex physical test setup.
Ker\"anen~\etal~\cite{kr-mbtes-12} investigate benefits of \ac{MBT} in HiL setups.
The employed \textit{online} \ac{MBT} approach generates test input on the fly and injects random test steps.
Randomness injection was also shown to be useful for testing interrupt-driven software by covering execution paths that are highly unlikely to occur~\cite{r-rtids-05}.
Combining \ac{MBT} and code generation was previously also demonstrated to support verification of model-based embedded systems~\cite{tksl-mbtmh-04}.
Downsides inherent to \ac{MBT} still apply, though: a correct model of the \ac{DUT} and additional development effort are needed, and vulnerability to human error exists.

Virzonis \etal~\cite{vjr-desuf-04} demonstrate how HiL, simulation, and CI can transform the linear development process of embedded control systems into iterative cycles.
Their work focuses on a process and local setup for the development phase whereas our work targets a distributed CI setup for the testing phase.

Muresan and Pitica~\cite{mp-slert-12} indicate that Software in the Loop is appropriate when a control algorithm is the test item.
In such cases, simulation benefits such as full parameter control and virtualization of timing critical aspects outweigh the downsides of not considering any hardware-related properties.

Common between the discussed HiL implementations is their usage in a very limited and well-defined domain for single specific purpose \acp{DUT}.
In contrast to that, we aim for a system that tests a wider range of general-purpose~devices.

\paragraph{Emulation and Simulation}
Embedded systems are sensitive to their physical environment making them cumbersome to test and debug.
Simulation and emulation solutions avoid these problems to make testing of embedded software easier for isolated aspects.
They promise benefits like reproducible experiments, major scalability improvements through parallel execution, faster operation by controlling virtual time, and many more.
QEMU~\cite{quemu} is a well-established machine emulator that provides many features for virtualization, user-mode emulation, up to full-system emulation.
Even though it can emulate constrained ARM MCUs as well as full-sized x86 and PowerPC systems it is more targeted towards the latter class of systems.
Renode~\cite{renode} is an open-source software development framework for running, debugging, and testing unmodified embedded software that is more focused on small embedded devices.
It employs simulation of the CPU in addition to its peripherals, externally connected sensors, and the communication medium between nodes to make multi-node system testing more reliable, scalable, and effective.

The benefit of such emulators is evident for cross-platform development and for providing an accurately controllable execution environment with a high degree of flexibility.
However, it is important to highlight that there are major obstacles that make it impossible to use these existing solutions for the same purpose targeted by our approach.
These obstacles mostly stem from two dimensions: the number of compatible devices and the basic applicability to test the behavior of low-level driver code with respect to the physical world.

RIOT as our test subject supports both aforementioned emulators~\cite{riot-emulators}, but none of them can provide the required level of compatibility.
To evaluate how far away current state-of-the-art emulation is from serving our intended purpose a look at the current support of devices and peripherals provides guidance.
According to the QEMU documentation, only two different embedded microcontrollers of the Cortex-M CPU family are supported\footnote{\url{https://wiki.qemu.org/Documentation/Platforms/ARM}}.
For Renode, the situation appears much better at first sight, with 9 targets being supported of our currently deployed 22 MCUs.
Though, as of now, feature availability puts up further limits with only two targets supporting all peripherals that our setup is testing.

Apart from existing solutions not being available for a relevant number of target devices there is also a fundamental mismatch of objectives compared to our solution.
Per definition, the objective of an emulator is to mimic or mock (\ie emulate) the behavior or a specific system.
While this covers the behavior of the (emulated) hardware towards the firmware, it does not cover the internal state of the real hardware and its interactions to the physical world.
In fact, existing emulators deliberately deviate internal system behavior to improve emulation performance~\cite{quemu-devices}.
Therefore, they lack the required level of simulation granularity to test for correct peripheral interactions with the physical world.
Albeit a hypothetical full-fledged \emph{simulator} would be capable of testing these aspects, to the best of our knowledge, currently no such simulator exists that covers both important aspects of device compatibility and basic applicability.
An accurate simulator will always require a high-quality model.
But the primary source to get detailed descriptions of hardware peripherals are manufacturer datasheets and reference manuals which differ considerably in terms of provided details and quality.
Additionally, they are subject to human interpretation and undiscovered hardware errata, arguably making it extremely challenging to derive exact models of hardware peripherals.
Implementing new device models would require non-negligible time whereas adding a new CPU with \philip only requires basic knowledge on how to provide peripheral configurations and wire the device.

We conclude that compared to emulation and simulation our approach has a conceptual advantage in terms of device compatibility, applicability to find errors in low-level driver code, and scalability when adding new devices.
With a scaleable way to qualify hardware behavior, our solution further contributes an important building block for future development and evaluation of generalized cross-platform MCU peripheral simulators.

\paragraph{Test Generation}
\label{testgen}
Research on test generation provided several methods to attain high coverage and unveil hard-to-find bugs with automated software testing.
Combinatorial, search-based, and \ac{MBT} allows optimizing test selection in order to slim down huge parameter spaces into manageable numbers of cases that still provide good coverage.
Symbolic execution is particularly interesting because of its ability to implicitly test huge sets of values by considering execution paths symbolically via constraints instead of concrete variable values.
A vast amount of tools is available from academia and industry that showed practical relevance and notable impact on automated software testing~\cite{cgkps-sestp-11}.
However, there are still multiple open problems related to \emph{path explosion}, \emph{path divergence} and \emph{complex constraints} that complicate employing symbolic execution for testing real production software at scale~\cite{abccc-osmas-13}.
For the constrained devices class we are interested in, symbolic execution was previously shown to be useful for testing firmware implementations of networked nodes~\cite{slawk-kdiib-10}.
Even though that approach can detect inter-node bugs in a network of embedded devices, it only considers network and application layer behavior, leaving MCU peripheral code out of scope.

In our specific domain --- testing peripheral drivers including their interactions with the physical world --- complex constraints, environment interactions, and device-specific characteristics can be expected to strongly interfere with symbolic execution, as is already visible with virtual devices~\cite{cxl-sevd-13}.
The approach developed by Cong~\etal~\cite{cxl-sevd-13} employs symbolic execution for virtual QEMU devices via the KLEE~\cite{cde-kuagh-08} engine.
It is evaluated with five different network devices and methods are proposed to reduce the manual implementation effort for the required device-specific models.
As indicated by the authors, however, the manual effort to enable symbolic execution for devices can not be completely avoided.
This leaves major blockers regarding applicability to our domain:
a widely applicable set of accurate software models for MCU peripheral devices does not exist, and
platform-specific hardware interfaces towards MCU-internal peripherals come in almost arbitrary variants making it very hard to automate model generation.

\section{Discussion, Conclusion, and Outlook}
\label{sec:discussion}

In this paper, we presented and evaluated a concept and implementation for HiL testing low-end IoT devices.
Part of the solution is two crucial components:
\philip, an open-source external reference device,
and a tool-assisted process that simplifies verification of peripheral behavior on different embedded hardware platforms.
In the following, we take a retrospective look at design decisions, report on lessons learned, and address shortcomings with potential solutions worth considering in future designs.

\subsection{Concept Validity}
The evaluation results show that \philip serves as a versatile tool to instrument currently 22 heterogeneous DUT devices with significant extensions planned in the near future.
Approximately 67\% of the peripheral implementation variants (I2C, SPI, UART) supported by RIOT OS are covered.
The timing analysis confirms the test throughput is high enough to increase the current $\approx 1,500$  nightly tests by an order of magnitude.
Future work could reduce the overhead of the test firmware on the DUT further by leveraging a more efficient serialization of commands and responses.

Employing the Memory Map Manager (\mmm) proved beneficial to simplify maintenance and enable resource-efficient operation in terms of processing time and memory.
The \mmm saves time compared to manually altering and reviewing the memory map when extending and maintaining \philip.
Instead of using a custom JSON-based configuration format, the memory map could in the future be encoded with the widely used SVD format~\cite{armsvd}.
This would allow using tools developed for SVD files alongside the \mmm.

Exposing low-level peripheral APIs via \ac{DUT} commands enables offloading test logic to the \tnode, keeping the size of test firmware static.
Offloading allows operation delays to stay below the order of \SI{100}{\milli\second}.
For time-critical operations, we showcased how asynchronous signal triggers of the DUT can be captured by \philip with a measurement precision on the microsecond scale.
Our practical case study during the I2C rework confirmed that \philip enabled better coverage of test cases and boards with less effort while ensuring easy to reproduce test setups and execution.
Results of \philip are continuously used by developers to identify and fix bugs.

The structured testing interface helped to evolve a multitude of customized test applications with unspecified architecture and coding styles into a uniform shape to build test suites that follow a consistent test design.
The developed wrapper tools provided machine-readable entry points that described available low-level APIs in a well-defined format.
This information can be used for higher-level test generation and orchestration tools for further automation of end-to-end testing~processes.

\subsection{Limitations}
\label{sec:lim}
\paragraph{MCU as a Reference}
We chose to implement \philip with an MCU (opposed to, \eg an FPGA) to simplify its design and make it accessible to embedded OS developers, though, an MCU has limitations.
The MCU must reply immediately when data is requested via peripheral buses, which cannot be guaranteed when the bus clock speed is high relative to the MCU clock speed.
Thus, advanced capabilities such as dynamic data response are provided only for lower transfer speeds, limiting high-speed transfers to predefined data.
Using an FPGA or a very fast MCU are potential solutions, however, they interfere with our original goal of being low-cost and accessible to typical embedded developers.

\paragraph{Static Wiring}
Most MCUs support several options to multiplex peripheral signals onto different pins or set pins to alternate functions.
In our current setup, peripherals are only tested using a fixed pin configuration based on the default pins provided by RIOT OS.
Tests are skipped if the wiring is not supported, for example, a peripheral is not exposed on the \ac{DUT}.
Even though this can validate peripheral behavior, it does not cover all possible deployment configurations.
While this can be addressed with a custom FPGA implementation or the addition of signal muxes, we argue that misconfigured pins are easier to find than software bugs.

\paragraph{Time-critical Test Commands}
Due to the command response-based interaction method, only synchronous commands that contain the communication overhead can be issued.
Wrapping time-critical interactions in firmware is a solution but loses the benefit of the agile process when many variations of wrapped commands are needed.

\paragraph{Capturing and Reporting Bugs}
Outside of the I2C rework case, the bugs that were produced in RIOT are not included as the focus was not on studying the bugs that were caught but providing reproducible tools for developers.
As of now the capabilities of \philip are still beyond the features used in the current test suites.
We prioritized maturity and reliability for developers over very sophisticated features to avoid false positives and support community adoption.

\paragraph{Coverage Feedback}
Code coverage metrics serve as valuable feedback to measure test completeness and for controlling test case generation.
Static analysis can be used for preliminary offline coverage assessment but the runtime dynamics of peripheral interactions render such approaches incomplete because external events (like an altered signal) affect the paths executed in low-level peripheral driver code.
Covering this requires on-target execution and end-to-end testing of firmware  with the involved hardware peripherals, as proposed in this paper.
In our setup, MCU-embedded trace macrocells or external debuggers can be employed for collecting runtime information on the \tnode to deduce code coverage.
A generic implementation of this is considered future work as platform-specific debugging interfaces need dedicated tooling and considerable integration work without providing intrinsic benefits on its own.
However, combining this with approaches for automated test generation is expected to be worthwhile the effort and mandates for separate examination.

\subsection{Stability}
\paragraph{Board Quality Issues}
A low-cost board such as the bluepill may introduce quality issues, especially when purchasing from different vendors.
The quality of the oscillator used for the clock source varies, affecting the minimum jitter and overall accuracy for timing tests.
This still allows us to find misconfigured clock trees.
Some calibration can be done to improve long-term drift tests.
The infrastructure tests and qualification procedures are also able to reduce quality issues.

\paragraph{Reliable DUT Interface}
Some vendor implementations of the USB interface used for communication between the \tnode and the DUT are not reliable, causing occasional test failures.
To improve the reliability we add retries which increased the setup time but handled spurious failures.

\paragraph{Flashing Problems}
Development tools for flashing DUTs are sometimes only available as closed-source software.
This either restricts the \tnode to a specific CPU architecture and OS or requires manual integration effort.
Even if flashers are readily available, many have reliability issues and cause occasional device lockups.
To cope with locked-up DUTs, \philip was equipped with automated flasher recovery mechanisms.

\subsection{Outlook}
In future work, we will use \philip and the process described in this paper to improve test coverage on an increasing amount of platforms.
To use \philip with other \acp{OS}, its tooling can be supplemented with additional implementations of protocol abstraction layers that wrap DUT interactions.
Information derived from regular HiL testing results and benchmarks of evolving software versions can be incorporated in various ways.
It provides valuable feedback for well-informed technical decisions and strategic planning of development efforts and gives users a comprehensive and realistic overview of the quality and performance of supported features across available target devices.
The system can further be used to qualify peripheral implementations of simulators and emulators by verifying them against the ground truth of a real device in the HiL setup.
With the instrumented hardware and automatic test execution in place, another promising direction is to equip our system with methods for coverage feedback and automatic test generation.

\begin{acks}
We are grateful to Pekka Nikander, who motivated much of this work, and Aiman Ismail, who implemented the timer test suite.  
We would like to thank the anonymous reviewers for their valuable feedback.
This work was partly supported by the German Federal Ministry of Education and Research (BMBF) within the project \emph{RAPstore} and the Free and Hanseatic City of Hamburg within \emph{ahoi.digital}.
\end{acks}

\section*{A Note on Reproducibility}
We explicitly support reproducible research~\cite{swgsc-terrc-17,acmrep}.
The source code and documentation of our designs and implementations (including tools and scripts to set up the testing) are available on Github, see \url{https://philip.riot-apps.net/}.

\bibliographystyle{acm}
\bibliography{local,systest,internet,iot,own,meta}

\end{document}